\documentstyle [11pt]{article}
\begin{document}

\begin{center}
{\Large\bf
N-body Simulations of the Small Magellanic Cloud and
the Magellanic Stream}
\vspace{1cm}

{\bf Lance T. Gardiner}

Department of Physics, Sun Moon University,

Tangjeongmyeon, Asankun, Chung-nam, Republic of Korea, 337-840.

\vspace{0.5cm}

{\bf Masafumi Noguchi}

Astronomical Institute, Tohoku University,

Aoba, Sendai 980, Japan.

\vspace{1.5cm}
{\bf ABSTRACT}

\end{center}

An extensive set of N-body simulations has been carried out on
the gravitational interaction of the Small Magellanic Cloud (SMC)
with the Galaxy and the Large Magellanic Cloud (LMC).
The SMC is assumed to have been a barred galaxy
with a disc-to-halo mass ratio of unity before interaction
and modelled by a large number of self-gravitating particles,
whereas the Galaxy and LMC have been represented by rigid spherical
potentials.
Our more advanced  numerical treatment has enabled us to
obtain the most integrated and systematic
understanding to date of numerous morphological and kinematical
features observed in the Magellanic system (excluding
the LMC), which have been dealt with more or less separately in previous
studies.
The best model we have found succeeded in reproducing
the Magellanic Stream (MS) as a tidal plume created by
the SMC-LMC-Galaxy close encounter
1.5 Gyr ago.
At the same time, we see
the formation of a leading counterpart
to the Magellanic Stream (the leading arm),  on the
opposite side of the Magellanic Clouds to the Stream, which
mimicks the overall distribution of several neutral
hydrogen clumps observed in the corresponding region of the sky.
A close encounter with the LMC 0.2 Gyr ago created another tidal tail and
bridge system, which constitutes the inter-Cloud region in our model.
The elongation of the SMC bar along the line-of-sight direction
suggested by Cepheid observations
has been partially reproduced, alongside its projected appearance
on the sky. The model successfully explains some major trends in
the kinematics of young populations in the SMC bar and older
populations in the `halo' of the SMC, as well as the
overall velocity pattern for the gas, young stars, and carbon stars
in the inter-Cloud region.

\vspace{0.5cm}
\noindent{\bf Key words:}galaxies:Magellanic Clouds - galaxies:interactions -
galaxies:kinematics
and dynamics - galaxies:structure -  methods:numerical

\newpage
\vspace{1cm}
\noindent
{\bf 1 \hspace{0.5cm} INTRODUCTION}

The interaction between the Large and Small  Magellanic Clouds (LMC
and SMC) and that between the Clouds and the Galaxy is believed to
have markedly influenced the evolutionary development of the Magellanic Clouds
 as galaxies. These interactions are thought to have produced several
observed tidal features such as the Magellanic Stream (MS) and inter-Cloud
region (ICR), and distorted
the internal structures of the Clouds such as can be seen in the large
depth of the SMC (e.g., Caldwell \& Coulson 1986) and shell-type features
in the stellar distribution of the LMC (Irwin 1991).

Renewed interest in the origin of the MS (see recent papers
by Moore \& Davis 1994, Sofue 1994, for example)
and the continued accumulation of observational data related
to the gravitational interaction of the Magellanic Clouds
(see Kunkel, Demers \& Irwin 1994; Hatzidimitriou, Cannon \& Hawkins 1993)
have motivated the development of a more advanced model to elucidate
the mechanisms ultimately responsible for these features.
We have built on the
results of the paper by Gardiner, Sawa \& Fujimoto (1994), hereafter Paper
I,  which was in turn based on the model of Murai \& Fujimoto (1980), to
develop a self-gravitating model of the Magellanic
Clouds system.
The model we present here is found to
match the recent observations by Kunkel, Demers \& Irwin (1994)
and Hatzidimitriou, Cannon \& Hawkins (1993), and also provides novel
explanations for other data related to the Magellanic system.\\

Viable orbits for the Magellanic Clouds about a Galaxy with an
extended massive halo were obtained in Paper I and are employed in this
present work. The inclination and orbital sense of these Magellanic orbits
agree with those
determined from proper motion observations of the LMC by Jones, Klemola \& Lin
(1994). Although the magnitude of the galactocentric transverse velocity of the
LMC derived in Paper I (287 kms$^{-1}$) is larger than that determined by Jones
{\it et al.} ($215\pm48$ kms$^{-1}$), subsequent work by Lin, Jones \& Klemola
(1995) employing their LMC transverse velocity measurement in conjunction with
various
Galactic potential models leads to a very similar picture for the
LMC orbit to that of Paper I.
Both Lin {\it et al.} and the authors of Paper I derived orbits for the LMC
with peri- and apo-galactic distances of about 45 and 120 kpc respectively, and
thus the interaction dynamics of our simulations are consistent with recent
observations.

The simulation of Paper I was able to
reproduce the basic observed features of the
MS and the ICR, and partially
succeeded in determining some
effects  of the interaction between the Clouds on their internal structure
by employing a test particle simulation.  Some limitations, however,
are encountered in representing the distribution of matter in the
Magellanic  Clouds by a sytem of massless particles orbiting in fixed
potentials.
Chief of these is that the alteration of the form
of the potential field due to the deformation of the internal  structure
of both the Clouds is not taken into account.
Secondly, the role of stellar bars in both the LMC and SMC  is ignored in
the test particle  simulation in which the potential is axisymmetric.
These considerations led us to develop a self-gravitating
model of the Magellanic system in which the total potential field at
a point may be derived from the summation  of the
forces due to all the constituent particles. As a first
application  of our self-gravitating model we present simulations
of the SMC and its associated tidal products, which include
the MS and ICR.

In the following section we present a description of the overall numerical
scheme for the system, including the adopted model parameters, and summarise
the main features of an equilibrium model for the SMC  generated
in preparation for the simulation of the Magellanic system.
In Section 3 we discuss the overall properties of our simulations
and how we searched for a model which best matches the
existing observational data.
Section 4 is devoted to the description of the best model
we obtained.
Finally, in Section 5 we state our conclusions
and discuss some future applications of our model.\\

\noindent
{\large\bf 2 \hspace{0.5cm} THE NUMERICAL METHOD}

Our computational model incorporates the basic framework of the model
of the Galaxy-Magellanic Clouds system described in Paper I, but instead of
representing each of the Magellanic Clouds as a test particle disc
in a rigid potential, here we model
one of the Clouds by a system of self-gravitating particles.
First, we review the general formulation of the dynamics of
the Galaxy-Magellanic
Clouds system.
Then we describe the {\it N}-body model used as the initial condition
for the SMC.

\noindent
{\bf 2.1 \hspace{0.5cm} The model of the Galaxy-Magellanic Clouds system}

In Paper I the model of the Galaxy, LMC and SMC, originally devised
by Murai \& Fujimoto (1980), was adopted with more recently derived
observational parameters in order to reproduce the global distribution of
matter in the Magellanic system. The Galaxy potential used was that
due to a spherical mass distribution possessing a  flat rotation curve
out to more than 200 kpc from the Galactic Centre. The LMC and SMC
were represented by Plummer-type potentials. The Magellanic Clouds were
considered  to experience a dynamical friction force due to their motion
through the dark halo of our Galaxy. Numerous test particle
computations were carried out which enabled suitable orbital parameters
for the Magellanic Clouds to be determined by modelling the MS
and inter-Cloud region. In Table 1 we summarise the observational parameters
used in the
model and list the estimated current space velocities of the Magellanic
Clouds obtained from this modelling. Additional model parameters to
be employed in the present work, and which are discussed in the
following subsection, are also tabulated.

Our present numerical scheme
differs from the simulations of Paper I in that we do not calculate the
evolution of both Clouds simultaneously. Instead we perform separate
computations
for each of the Magellanic Clouds, constituting one of the Clouds
as a self-gravitating particle system and representing
the gravitational influence of the Galaxy and the other Cloud by
fixed potentials identical to those used in Paper I.
Hereafter we denote one of the Magellanic Clouds, that which is represented
by a particle system, as `Cloud 1', and the Cloud represented by the fixed
potential as `Cloud 2'.
Our procedure is described as follows:\\
1. We construct an equilibrium model for Cloud 1 (the `equilibrium run', see
Section 2.2) and
subsequently place it at the origin of a non-inertial Cartesian coordinate
system centered on Cloud 1. \\
2. The orthogonal axes of this non-inertial coordinate
system are oriented such that the disc of Cloud 1 is coincident with
the $x-y$ plane and the $z$ axis coincides with the rotation
axis (the axis of the spin angular momentum vector) of Cloud 1. In the general
case, the
orientation of the non-inertial
system relative to the inertial (galactocentric) system is specified
by the direction of the rotation axis
with respect to the galactocentric system ($X,Y,Z$). Two angles, $\theta$
and $\phi$, defined  in the spherical polar coordinate system in the
usual way (see Fig.1),
determine the relationship between the two systems.\\
3. Based on  the orbits derived in Paper I for the time interval from
$T=-2$ Gyr to $T=0$ which corresponds to the current epoch,
we create a look-up table of positions and velocities for the Galaxy
and Cloud 2 in our non-inertial system ($x,y,z$).
The timestep
used has a duration of $2\times10^6$ yr giving 1000 timesteps for the
complete simulation.\\
4. Cloud 1 is evolved from $T=-2$ Gyr
towards the present under the influence of external forces due to the Galaxy
and Cloud 2 (the `interaction run'). \\
5. In the present paper we take the SMC to be Cloud 1 and the LMC to be
Cloud 2. The particles in Cloud 1 are assumed to be collisionless.
Then
the complete expression for the total force applied
to an individual particle in Cloud 1 is given by:\\

$$ \ddot{\bf r}_i = -{{G} \over {m_i}}
 \sum_{j \neq i}^{n} {{m_j({\bf r}_i-{\bf r}_j)}
  \over {(|{\bf r}_i-{\bf r}_j|^2+\epsilon^2)^{3/2}}}
  +{\bf F}_{Gal}({\bf r}_i-{\bf r}_{Gal})
  +{\bf F}_{lmc}({\bf r}_i-{\bf r}_{lmc})
  -{\bf F}_{Gal}(-{\bf r}_{Gal})
  -{\bf F}_{lmc}(-{\bf r}_{lmc})
   , \eqno(1) $$

\noindent{where ${\bf r}_i$ denotes the position vector of the $i$-th
particle in the coordinate system ($x,y,z$) centered on Cloud 1,
$m_i$ is the particle mass, and $G$ is the gravitational constant.}
The softening parameter is denoted by $\epsilon$ and is discussed later.
The positions of the Galaxy and the LMC
in this coordinate system are denoted by ${\bf r}_{Gal}$
and ${\bf r}_{lmc}$,    respectively, and these should be understood as given
functions
of time. We assume that the gravitational potential of the Galaxy
is one which gives a flat rotation curve with a constant circular
velocity, $V_0$ (=220 kms$^{-1}$), out to beyond 200 kpc from the Galactic
Centre
so that the gravitational force of the Galaxy exerted on
a particle of unit mass is given by

$$ {\bf F}_{Gal}({\bf r})=- {{V_0^2} \over {|{\bf r}|^2}} {\bf r}. \eqno(2)$$

The LMC is assumed to have a Plummer-type potential
with an effective radius, $K$ (=3 kpc), giving a
gravitational force on unit mass of

$$ {\bf F}_{lmc}({\bf r})=-{{GM_{lmc}{\bf r}} \over {({\bf r}^2+K^2)^{3/2}}},
\eqno(3) $$

\noindent{where $M_{lmc}$ is the total mass of the LMC and taken to be
$2 \times 10^{10} M_{\odot}$.}
The last two terms in equation (1) are the
correction terms which arise from  integrating the equations of motion
in a {\em non-inertial} coordinate system centered on
the SMC. It should be noted that in the present formalism the effects of
dynamical friction are implicitly included in the time development of
${\bf r}_{Gal}$ and ${\bf r}_{lmc}$.

\vspace{1cm}
\begin{center}
------

Fig. 1

------
\end{center}
\vspace{1cm}

\noindent
{\bf 2.2 \hspace{0.5cm} Equilibrium model for the SMC}

As discussed in the previous subsection, we first constructed an equilibrium
model for the SMC as Cloud 1. Here we discuss some choices
in the physical parameters adopted for the equilibrium model.

\noindent{\it
a) Global structure of the SMC}

We constructed an equilibrium model for the SMC as a two-component
system consisting of a nearly spherical halo and a rotationally supported disc.
 Apart from the fact that such a two-component system represents a good
approximation of a real disc galaxy, it facilitates the comparison of the
numerical results with the observational data for objects of various
age groups such as horizontal-branch stars (belonging to the halo)
and Cepheids (belonging to the disc).

Both the halo and the disc are truncated at a radius of 5 kpc
(the disc radius).  The tidal radius of the SMC, $r_t$, is given by

$$ r_t=r_p [{{M_{smc}} \over {(3+e)M_{Gal}}}]^{1/3} \eqno(4)$$

(Faber \& Lin 1983), where $r_p$ is the perigalactic distance of the SMC,
$M_{smc}$ is the mass of the SMC,  and $M_{Gal}$ is the mass of the Galaxy
contained
within $r_p$.
Here the orbital eccentricity, $e$, is given by
$e=(1-b^2/a^2)^{0.5}$, where $a$ and $b$ are the semimajor and semiminor
axes of a pseudo-ellipse, respectively.
Note that the orbit is not an exact ellipse because the Galaxy potential
is not that due to a point mass.
The orbital data give
values of 43 kpc and 0.89 for  $r_p$ and $e$ respectively. The formula
yields a tidal radius of 5 kpc for the SMC, which is equivalent to the
adopted   truncation radius.

In this present work we use an SMC
mass of $3\times10^9 M_\odot$, compared to $2\times10^9 M_\odot$, used in
Paper I.
If we assume that the MS and ICR originate from the disc component of the SMC
and consider the gas associated with the SMC itself to belong to the disc
component, we have a combined mass of about
$1.3\times10^9$ for the original disc gas (Westerlund 1990).
In addition, the stellar disc component may also contribute
to the total disc mass making the original disc mass of the SMC
significantly
greater than $1.3\times10^9 M_{\odot}$.
Evidence for the existence of a halo
component in the SMC comes from
the spheroidal kinematics of carbon stars (Hardy, Suntzeff \& Azzopardi
1989) and planetary nebulae (Dopita {\it et al.} 1985).
Regrettably, no reliable estimate for the halo mass (possibly
including invisible matter) out to the adopted
disc radius is available at the moment. If we take a disc-to-halo mass
ratio of 1:1, then the total mass of the SMC in the pre-encounter
phase is at least $2.6\times 10^9 M_{\odot}$. This consideration
led us to take a relatively large mass for the SMC.
It should be pointed out that, due to the
much smaller mass of the SMC relative to the LMC and the Galaxy, the
SMC's orbital motion is nearly independent of the SMC mass.

The global structure and evolution of a model disc galaxy are largely
determined by
two  non-dimensional quantities, the disc-to-halo mass ratio and the turnover
radius
of the rotation curve relative to the disc radius.
The mass ratio between the disc and halo components can dramatically alter the
evolution of a system in an
isolated state. A stellar disc having a mass comparable to or larger
than that of the halo is known to develop a bar in a few rotations (e.g.,
Ostriker and Peebles 1973). Unfortunately, a reliable estimate of this
important parameter for the SMC {\it in the pre-encounter state} is
quite difficult to obtain.
We have chosen to use a disc/halo mass ratio of 1:1, because a number of
disc galaxies with reliable observational data have been shown to have
a mass ratio of around unity (e.g., van der Kruit \& Searle 1982).
It should be borne in mind that the  possibility of a smaller mass ratio,
and hence a pre-encounter disc stable against spontaneous bar formation,
cannot be ruled out. We assume that
the surface density distribution of the disc obeys an exponential law
with a scale length of 0.25 times the disc radius as
suggested for most disc galaxies (e.g., Fall 1980).

Another important parameter characterising the equilibrium model
is the turnover radius of the rotation curve. This is the radius at which
the rotational velocity changes from that of nearly rigid rotation to
a nearly constant value.
Numerical studies (e.g., Sellwood 1981) indicate that the length of the
spontaneously induced bar has a strong positive  correlation
with the turnover radius.
The turnover radius has been set to
a relatively large value of 3.5 kpc (i.e., 70 percent of
the disc radius) since a galaxy of the
Magellanic type (i.e., a low-luminosity galaxy of very late morphological type)
 generally has a slowly rising rotation curve from the center to the edge
(e.g., Rubin {\it et al.} 1985). These choices in the adopted parameters
allow
the generation of a stable bar $\sim 5$ kpc long  in the
isolated model as we see below.

\noindent{\it
b) Equilibrium run}

We initially distributed 10000 particles in the disc component and
5000 particles in the halo since we are mainly interested
in the disc particles, which, due to their smaller random motions,
tend to form finer structures than the halo particles.
This means that
in our calculations the mass of an individual halo particle is
twice that of a disc particle given a disc/halo mass ratio of unity.
In the calculation of the gravitational force we used the tree-code (e.g.,
Barnes and Hut 1986). The gravitational softening length ,$\epsilon$
(see equation (1)), to suppress
undesirable two-body effects was taken to be 50 pc.

The equilibrium run was divided into two stages.
First, only the halo particle system was evolved with the disc particle
system fixed. During this stage,
the halo particles experience the gravitational force of the disc
in addition to that of the halo itself, but the disc
component does not respond to the change of the total gravitational field.
The halo was found to reach a near-equilibrium state after several dynamical
times.
After this we calculated the gravitational force
(arising from both the halo and the disc components)
acting on each disc particle.
Each disc particle
was then given a rotational velocity and a small random motion which
corresponds  to the Toomre's (1964) $Q$ value of 1.5 so that
in the absence of any instability or external perturbation it would move on a
nearly circular orbit in balance with the gravitational force.
The resultant rotation curve is
slightly different from the initially specified one due to a small
change in
the halo density distribution.

Then the second stage was performed in which all the halo and disc particles
were moved
under their own gravity.
It was found that the disc quickly forms a bar structure within two disc
rotation  periods in accordance with many previous studies. After having fully
developed,
the bar did not significantly alter its shape or angular velocity
(pattern velocity). Therefore we adopted the state
after about 4 rotation periods
as the initial condition for the interaction simulations
described in the following.
Fig.2 plots the spatial distribution of the disc and halo particles separately
for this initial state.
The halo is nearly spherical whereas the disc develops a strong bar
with a length of $\sim 5$ kpc.
Non-circular motions along elongated orbits in the disc plane dominate
the disc after the bar has fully developed, whereas the halo is mainly
supported by random motions.\\

\vspace{1cm}
\begin{center}
------

Fig. 2

------
\end{center}
\vspace{1cm}

\noindent
{\large\bf 3 \hspace{0.5cm} INTERACTION  SIMULATIONS}

In the interaction runs, the SMC model is evolved under the influence of
external forces due to the Galaxy and the LMC, starting from the initial
condition mentioned above.

\noindent
{\bf 3.1 \hspace{0.5cm} General trends}

The first important step was to determine the spatial orientation
angle ($\theta, \phi$) of the SMC disc.
If the disc orientation is changed, the orbits of the Galaxy and the LMC
in the SMC-centric coordinate system should be correspondingly adjusted.
We decided to use a different
initial spatial orientation for the  disc of the  SMC from that used in
the test particle computations of Paper I. Owing to the overall irregular
structure and highly disturbed internal kinematics of the SMC,
the correct orientation to use
is far from clear. In Paper I the authors adopted a disc orientation
parallel
to the galactocentric $X-Z$ plane consistent with de Vaucouleurs' (1960)
determination
of the inclination and major axis of the SMC from star counts.
(This orientation is specified by $\theta=90^\circ$, $\phi=270^\circ$
in our system.)
However, with this orientation the SMC bar would lie parallel to the
sky plane, but according to the observations of Caldwell \&
Coulson (1986), we require the bar to be mainly oriented along our
line-of-sight in its final position.
To satisfy this requirement for the bar, as well as the
requirement that the major axis of the disc be oriented along the
observed major axis, we adopted a disc spatial orientation specified
by  $\theta=45^\circ, \phi=230^\circ$.
It turned out that the bar is located nearly in the original
disc plane at the present epoch in all the models calculated,
and thus
it was possible with this disc orientation to
produce simulation results in which the bar is oriented mainly along
our line-of-sight.
On the other hand, we could not construct any satisfactory models
by adopting $\theta=90^\circ$, $\phi=270^\circ$ (see Section 4.2).
It is to be noted that the adopted values of
$\theta=45^\circ, \phi=230^\circ$ may not comprise a unique solution but
could be one of a number of possible space orientations.

The final orientation of the bar with
respect to our line-of-sight will depend also on the initial
orientation of the bar within the SMC disc plane (i.e., the $x-y$ plane
of the SMC centric coordinate system). Hereafter, the angle between
the bar major axis (at the beginning of an interaction run)
and some defined axis in the $x-y$ plane is referred to
as the bar position angle.
This angle is a
free parameter in our simulations. In order to achieve the best fit
of the bar spatial orientation to the observations we therefore generated
a series of models with different initial bar postion angles
(keeping $\theta=45^\circ $ and $\phi=230^\circ$).
We have carried out
12 simulations for bar angles
separated by an interval of $30^\circ$. From now on we denote a model with its
three parameters, $\theta, \phi$
and $p$, as $\theta / \phi /p$, where $p$ is the initial position angle
of the bar. For example, the model 45/230/180 has $\theta=45^\circ,
\phi=230^\circ$, and $p=180^\circ$.
For each of our 12 models with different initial bar position angles, we
constructed a series of plots to identify the model which best fits
the observational properties of the Magellanic system. The plots
constructed were based on Figs 6, 7 and 13 from Paper I.
Using these plots, we looked for good geometrical structure in the
simulated
Magellanic Stream, and good
agreement of the spatial orientation of the bar and the appearance of
the SMC projected on the sky plane with the observations.
It was found that the morphology of the MS  displayed a satisfactory agreement
with the observed geometry and was largely independent
of the initial bar position angle. Therefore the main consideration
which determined the choice of our best model was the
agreement of the spatial orientation of the bar with observations. All models
produced a well developed
leading arm on the opposite side of the MS, which is discussed in
detail in Section 4.2.
Pairs of models with their initial bar position angles differing by $180^\circ$
were found to have similar bar orientation at the present epoch,
reflecting
the nearly bi-symmetrical nature of the initial barred model.
A bar orientation which matched the observations was
best reproduced for $p = 270^\circ$ or $ 90^\circ$ and it was decided
to adopt model 45/230/270 as the best model.

\noindent
{\bf 3.2 \hspace{0.5cm} Interaction dynamics}

Detailed discussion of the best model is postponed to the next section.
Nevertheless, it is worthwhile to mention here some aspects of the
SMC-LMC-Galaxy interaction in the best model.
Fig.3 shows the orbits of the Galaxy and LMC around the SMC in the
SMC-centric coordinate system ($\theta=45^\circ$, $\phi=230^\circ$)
along with the time evolution of
the separation between the SMC and the Galaxy or the LMC.
The positions corresponding to the times at which the LMC-SMC separation takes
a local minimum are marked
on the orbits. The SMC disc rotates counterclockwise in the $x-y$ plane.
It is seen that both the Galaxy and LMC orbits are roughly polar.
The relative orientation of the SMC disc is thus not the most
favourable for tidal distortion but we can still expect a significant
tidal effect when the Galaxy and/or the LMC pass by the SMC at a small
distance.
Although the Galaxy-SMC separation always stays larger than the LMC-SMC
separation, the more relevant quantity here is the strength of the tidal
force,
which is depicted in Fig.4.
The tidal force in Fig.4 has been calculated by differentiating the
gravitational potential of the indicated galaxy twice. In other words,
we calculated the gradient of the gravitational force. This gradient
gives the relative strength of the tidal force if the perturbed body
(the SMC in this case) has a constant size.
Although this condition is not satisfied exactly in
practice, Fig.4 is still instructive.
It is clear that at $T \sim -1.5$ Gyr, when both the
Galaxy and the LMC pass the peri-SMC points, the tidal forces due to the
Galaxy and the LMC are comparable in magnitude. The MS starts to develop
roughly at this epoch. It is therefore suggested that the early development
of the MS is governed by the combined effect of the Galaxy and the LMC.
At $T \sim -0.2$ Gyr, both the Galaxy and LMC make second close encounters
with the SMC. The LMC tidal force is much stronger than that of the
Galaxy at this epoch,
and is thought to have played a major role in the shaping of
tidal features in this most recent epoch.

\vspace{1cm}
\begin{center}
------

Fig. 3

------
\end{center}
\vspace{1cm}
\vspace{1cm}
\begin{center}
------

Fig. 4

------
\end{center}
\vspace{1cm}

\noindent
{\large\bf 4 \hspace{0.5cm} THE BEST MODEL --- COMPARISON WITH OBSERVATIONS}

\noindent
{\bf 4.1 \hspace{0.5cm} Global features of the model}

In this section we discuss the simulation results for
the best model, with parameters 45/230/270.
We first show plots of the global structure produced by the simulation
to define our terminology.
In Fig.5(a,b) we have plotted, for both disc and halo components,
the particle positions in the SMC-centric $x-y$ plane at $T=0$,
namely the current epoch.
The various features produced by the interaction of the SMC with
the Galaxy and LMC are labelled on the plot for the disc component
(Fig.5a), namely the Magellanic Stream,
leading arm, and tidal bridge and tail. The tidal bridge and tail were
generated by the close encounter between the Magellanic Clouds about 0.2 Gyr
ago, while the MS and the leading arm originated at an earlier epoch
corresponding to the time of the previous perigalactic approach of the SMC,
and coincidentally the time of another close encounter between the
Magellanic
Clouds about 1.5 Gyr ago. Examination of the best
model (see section 4.3) reveals  that the tidal bridge and tail
are seen to a large extent overlapped in the sky, with the tail section more
distant than the bridge, thus giving rise
to the inter-Cloud region.
The halo component plot (Fig.5b) shows several differences from the disc
component plot, including a much less conspicuous stream at the the
position of the MS and a less well-defined leading arm.

Before embarking on a detailed discussion of the simulation results
in the following subsections, we here present an overview of
the relationship between our best model and the results of previous
simulations by other workers.
Previous simulations of the Magellanic Clouds system have largely aimed
at reproducing the geometry and kinematics of the MS,
with the notable exceptions of those produced by the authors of Paper I and
by Kunkel, Demers \& Irwin (1994) (KDI), in which the emphasis was on
the internal structure of the Magellanic Clouds.
A consensus emerged in the early 1980s on the basic type of
{\em tidal} model required
to realistically simulate the observational characteristics of the MS.
(Other models for the MS based on ram pressure stripping have
also been proposed --- see e.g., Moore \& Davis 1994, Sofue 1994.)
The simulations of Lin \& Lynden-Bell (1982) and Murai \& Fujimoto
(1980) (MF) settled on models in which the Galaxy has an extended massive
halo and
the Magellanic Clouds have polar orbits with the Clouds leading the MS.
The dynamics
of the Galaxy-LMC-SMC system in our model are fundamentally similar to
those of MF, and therefore our present model of the MS can
largely be considered as a refinement to the existing picture.
We do, however, present a more convincing interpretation
of the leading arm feature than in previous work (see next subsection).
Although two sets of models by Fujimoto \& Sofue (1976, 1977) have claimed
to reproduce the leading arm as tidal debris from the LMC and SMC, these
models could not simultaneously
achieve a good reproduction of the MS.
One of the models by Tanaka (1981) (see his Fig.5c) gives a fairly good
reproduction
of both the MS and the leading arm as tidal debris from the SMC,
but the radial velocity of the tip of the MS is more
positive than observed and the resulting leading arm (this is the tidal
tail in his model in which the Magellanic orbits are of opposite sense
with respect to our model)
forms a narrow line unlike the scattered distribution of
the real H~{\sc i} clouds.

A number of previous models have considered the internal structure
of the Magellanic Clouds by treating the LMC and SMC as separate entities.
This approach contrasts with one in which the Clouds are
considered as a single-entity (e.g., Lin \& Lynden-Bell 1982).
MF simulated the region between the Magellanic Clouds (ICR),
and their simulation also indicated that the SMC was greatly extended
along the line-of-sight direction. In Paper I the authors used a much larger
number of particles in their test particle simulation than in MF
and treated the structure of the
Magellanic Clouds in greater detail. The key feature of their model
was the generation of a tidal bridge and tail system which
could qualitatively reproduce the inter-Cloud region and the large
line-of-sight extension of the SMC. In our $N-$body simulation based
on the same orbital dynamics as Paper I we have also generated similar
structures.

The model of KDI  also generated a bridge-tail structure with several
similar features to the present model. The basic dynamics of their
model are different from ours in that they have neglected the gravitational
influence of the Galaxy and have adopted an unbound orbit of the SMC
with respect to the LMC.
Thus the formation of the MS was inevitably excluded from their
discussion.
Their simulations
have led to a different
interpretation of the kinematics of the inter-Cloud region from ours, but
nevertheless
their model reproduces some aspects of the structure and kinematics in the
eastern part of the SMC and the ICR
(see Section 4.4).

To summarise, the global structure of our model has much in common
with previous models, but is the first to {\em simultaneously}
explain many structural and kinematical features of the Magellanic system in
a single model. The greater sophistication of our model, which includes
particle self-gravity and a two-component disc/halo system representing
the SMC, enables it
to address a wider range of observational data than previous simulations.
We now analyse in detail the simulation of different aspects of the
Magellanic system by dealing with the MS, the SMC and the ICR in turn. We have
used a combination of
techniques adopted in Paper I and newly developed methods
to compare the simulation results with observations.

\vspace{1cm}
\begin{center}
------

Fig. 5

------
\end{center}
\vspace{1cm}

\noindent
{\bf 4.2 \hspace{0.5cm} The Magellanic Stream}

The MS is a narrow band of neutral hydrogen emerging
from near the Magellanic Clouds and extending for more than 100 degrees
in the sky.  It is generally believed that it is a tidal feature produced
as a result of the interaction between the Magellanic Clouds and the Galaxy.
The first successful attempts to model the MS as tidal
debris torn from the Magellanic Clouds were those of
Murai \& Fujimoto (1980) and Lin \& Lynden-Bell (1982).
These investigators surmounted the problem of achieving high
negative radial velocities at the tip of the MS by introducing
a Galaxy with a massive halo. In Paper I some major
characteristics of the MS were also obtained in a reproduction that is
fundamentally  similar to that of Murai \& Fujimoto (1980). The geometrical
appearance of the simulated Stream projected on the sky, however, was not
as satisfactory as one might  have hoped, and therefore
in this first application
of the self-gravitational model of the Magellanic system we sought to
achieve rather better agreement with the observational features
of the MS.
Although fundamentally the collisionless disc particles of our simulation
should represent the {\it stellar} disc component, it can be considered that
in areas of low particle density such as the MS,
the collisional and dissipative nature of the gas
is not strongly manifested and therefore a collisionless model is adequate for
modelling the gaseous MS.

\noindent
{\em a) Geometry of the Magellanic Stream}

In two main respects the model Stream in Paper I
failed to match the observations closely. First, the model Stream emerged
rather near the LMC on the plane of the sky, whereas in reality
it begins near the SMC.
Second, the model Stream was rather poorly
populated
when compared to the observations, especially at the start
of the Stream.
Our self-gravitating simulations for the spatial orientation
($\theta=90^\circ$, $\phi=270^\circ$),
corresponding to that used in the test particle simulation, similarly
produced a model Stream originating from near the LMC instead of
the SMC. However, for our best model (parameters 45/230/270) these
deficiencies are overcome, as
we will see shortly. In Fig.6 we show, for our best model,
a plot of the distribution of disc particles (right panel) compared
with the neutral hydrogen distribution (left panel) of the MS
projected onto the sky centred on the South Galactic Pole.
The plot of the distribution of neutral hydrogen shows data taken from
Mathewson \& Ford (1984) plus other observations of H~{\sc i} clumps on the
opposite side of the Magellanic Clouds from the MS. These latter  observations
made with coarser resolution were derived from Mathewson, Cleary \&  Murray
(1974) and the data are indicated by the thicker curves on the plot.

Using Fig.6, we see that a well defined stream of particles extending over
$\sim 100$ degrees in the plane of the sky emerges from the ICR
near the SMC,
in better agreement with the observed geometry of the MS.
There are approximately 1300 disc particles in the model Stream, which
corresponds to about $2\times10^8 M_\odot$ of material, of the same order
of magnitude as observational estimates.
The simulated Stream is seen to comprise two separate
streams, namely a more densely populated main stream which lies close to
the position of the actual MS, and a less conspicuous
secondary stream to its left.
The  secondary stream is not actually seen in the neutral hydrogen
observations,  suggesting some difficulty with the model.
This may, however, not be so serious,  because the expected
surface density in the secondary stream is much lower  compared
to the main stream.
Disregarding the secondary stream, the model MS is relatively broad
at its tip and its origin near the ICR.
We note that the actual MS shows similar structure.
Our success in achieving a more realistic reproduction of the observed
morphology
of the MS is a notable feature of the present model.

\noindent
{\em b) The velocity profile}

Fig.7 plots the radial velocity seen from the sun corrected for the
motion of the LSR with respect to the Galactic Center ($V_{GSR}$)
against the Magellanic longitude defined along the MS by Wannier and Wrixon
(1972). (The `true' galactocentric radial velocity is the velocity which would
be observed
from the Galactic Centre itself, and owing to the small offset of the sun's
position
with respect to the Galactic Centre there is a slight difference
between this quantity and $V_{GSR}$. For simplicity, hereafter
the `galactocentric radial velocity' is used to mean
the velocity seen from the sun but corrected for the solar rotation about the
Galactic Centre).
Observed velocities for the H~{\sc i} gas are denoted by large diamonds. The
model  shows  reasonable agreement with the observational data.
Both the high velocities at the
beginning of the MS, about 100 kms$^{-1}$ in the vicinity of the Magellanic
Clouds, and the high negative radial velocity of $-200$ kms$^{-1}$
at the tip of the MS are reproduced.

\noindent
{\em c) The leading arm feature}

Here we mention another interesting aspect of this best model.
In Fig.6 (simulation plot in the right panel) an inverse L-shaped leading arm
can be seen
on the opposite side of the Magellanic Clouds to the MS. This leading arm
is more sparsely populated with particles than the MS.
We consider that this feature corresponds to
the several H~{\sc i} clumps observed by Mathewson, Cleary \& Murray (1974)
in the area defined by
$260^\circ<l<310^\circ, -30^\circ<b<30^\circ$.
The observed H~{\sc i} clumps, though discrete,  delineate
a similar inverse-L shape whose position on the sky roughly agrees with the
model. The positional agreement is not so precise since the simulated `arm'
extends to galactic latitudes as high as $b=60^\circ$.
The galactocentric radial velocities of the particles in this simulated
leading arm, which range from 100 to 200 kms$^{-1}$, are
somewhat larger than the observed values for H~{\sc i} clumps, in the range
0 to 100 kms$^{-1}$, but both model particles and  H~{\sc i} clumps exhibit
a flatter trend in velocity with respect to Magellanic longitude contrasting
with the systematic decrease in velocity seen along the length of the MS. The
detailed velocity
profile and distribution of matter in the leading arm will depend on the
form of the LMC potential which appears to be responsible for
scattering material of SMC origin into its present location.

A globular cluster, Ruprecht 106, has been discussed by Lin \& Richer (1992)
as a possible candidate for an object that has been tidally captured from the
Magellanic Clouds
by the Galaxy. It is actually located in the leading arm region on the
sky. However, the observed galactocentric radial velocity of
$\sim -233$ kms$^{-1}$ is much lower than that of the H~{\sc i} clouds
observed by
Mathewson, Cleary \& Murray (1974) in this region, which have velocities
exceeding 0 kms$^{-1}$. Therefore, its association with the leading arm is
very doubtful. Irwin (1991) reports on four carbon stars near $RA=13^h,
Dec=0^\circ$. In galactic coordinates this corresponds to $l=310^\circ$,
$b=60^\circ$,  a location
close to the `elbow' of the leading arm. However, no velocity data is
available
for these carbon stars for comparison with our best model
or the Mathewson {\it et al.} (1974) observations. Thus the existence
of a stellar counterpart to the H~{\sc i} clouds
in the leading arm region has yet to be confirmed.

\vspace{1cm}
\begin{center}
------

Fig. 6

------
\end{center}
\vspace{1cm}

\vspace{1cm}
\begin{center}
------

Fig. 7

------
\end{center}
\vspace{1cm}

\noindent
{\bf 4.3 \hspace{0.5cm} The SMC -- Internal structure and kinematics}

In the test particle simulation of Paper I it was shown that the
existence of the gaseous and stellar bridge between the Magellanic Clouds and
the large
extent in depth of the SMC could be explained by the creation of
a tidal bridge and tail system as a result of the close encounter with the
LMC about 0.2 Gyr ago.
As we will see in more detail later, our best model also reproduces
a basic bridge and tail structure (see Fig.5), with some notable differences
from the
previous work which result in a better match with several observational
features. Since our simulations were carried out including both disc
and halo components it is profitable to compare these simulation data with
extensive observations of both Population I objects
and older populations such as carbon stars and
horizontal-branch/clump stars.

\noindent
{\em a) Appearance of the SMC in the sky plane}

We begin by presenting plots showing the neutral hydrogen distribution
in the vicinity of the LMC and SMC,
and the distributions of disc and halo particles in our best model
projected onto the sky plane (Fig.8).
For the disc component (middle panel, Fig.8), it is seen that the SMC bar
is oriented
in a NE-SW direction, coinciding well with the actual orientation
of the bar of the SMC on the sky.
The disc component shows a boundary
in the particle distribution to the SW of the SMC, contrasting with
the broad and extended distribution of particles to the east and to the
northern direction. This overall distribution is also seen in the
H~{\sc i} distribution, which displays a sharp edge in the SW, a bridge of
gas extending to the LMC and extensions towards the north which form the
beginnings of the MS. It should be pointed out that
the disc component does not reproduce the detailed distribution
of gas in the ICR, and that the particle
distribution between the SMC and LMC is not so sharply concentrated as in
the main ridge of the observed
gas distribution, which runs along a line of constant
declination given by $Dec=-74^\circ$.
Also, the edge of the distribution in the SW is not as sharp as in the
observed
H~{\sc i} distribution.
This lack of agreement with the detailed observations could be
due to our neglect of gas dissipation processes in
our collisionless model.

The halo component (bottom panel, Fig.8) shows a largely circular projected
distribution within about 5 degrees of the SMC centre and a distribution of
particles extending into the ICR. Gardiner \& Hatzidimitriou
(1992)
found that the projected distribution of horizontal-branch/clump stars, which
may be considered
to belong to the intermediate-aged population between 2-15 Gyr old,
was also largely circular at around distances of 5 kpc
from the SMC centre.
In their review
of the distributions of various population groups, Azzopardi \& Rebeirot (1991)
 pointed
out that the older stellar population groups such as carbon stars and planetary
 nebulae have a rounder and less centrally concentrated distribution than
the
younger population groups, which are concentrated in the bar. Thus,
our simulation of the disc and halo populations is in broad agreement
with this general picture. The recent
discovery of carbon stars in the ICR by Demers, Irwin \& Kunkel
(1993), suggests that there may well be stars belonging to the
intermediate-aged component displaced into the ICR by tidal forces
in agreement with the model.
We later  discuss the detailed velocity structure of these carbon stars
(see Section 4.4).

\vspace{1cm}
\begin{center}
------

Fig. 8

------
\end{center}
\vspace{1cm}

\noindent
{\em b) The 3D structure}

Our discussion of the three-dimensional geometrical structure of the
SMC is centred on Fig.9, in which we have
plotted the distance of the simulation particles from the sun against
the angle along the maximum gradient defined by
Caldwell and Coulson (1986) (CC), which is in position angle $58^\circ$
on the
sky and runs approximately from NE (negative angle) to SW (positive angle).
In the left panel we have plotted the simulation data for the disc component
with the Cepheid observations by CC superimposed, while the right panel
shows the simulation data for the halo component.
In plotting the Cepheid data, we simply assumed that the distance
modulus of the coordinate centre used by CC (i.e., $\alpha=0^h51^m,
 \delta=-73^{\circ}.1$ (1950)) was 18.78 (Feast \& Walker 1987) and
for each Cepheid we used
the difference in modulus from this centre value given in
Fig.7 of CC.

The Cepheid observations by CC revealed a large elongation of their
distribution along
the line of sight. Observations by several other authors
(e.g., Florsch, Marcout \& Fleck 1981,
Laney \& Stobie 1986, Mathewson, Ford \& Visvanathan 1988)  have confirmed
a large depth for the SMC and suggested an overall positive gradient in
distance from the NE to the SW end of the bar. The disc component in our
simulation
(Fig.9, left panel) shows a bar highly inclined to the sky plane
with the NE end nearer than the SW end in agreement with the general
observed trend. Although the spatial orientation of
the bar matches the observations, the  model seems to show significantly
less elongation than that indicated by the Cepheids, with a model bar 5 kpc
long compared to an observed bar about 10 kpc long.
This discrepancy may not be serious. Cepheids, being young objects, will
better trace the interstellar gas component both spatially and kinematically
than will older stellar populations.
In the model, the disc particles are treated as
collisionless, and hence can be regarded as representing the
relatively older stellar population of the disc
formed before the start of the simulation at $T=-2$  Gyr.
It would not be surprising if the stellar and gaseous components exhibited
different dynamical responses, especially in a heavily disturbed region
of high gas density.
Therefore, a satisfactory detailed modelling of the {\it internal} structure
of the SMC should await a gas-dynamical simulation.
It is also seen that the tidal bridge and tail emanate from
the far-side and near-side of the bar, respectively. The bridge runs towards
negative angle (i.e., to the NE) and continues into the inter-Cloud
region. The tail first protrudes to the SW and then turns back to the
NE region, extending into the ICR as well.
We notice that
the bridge and tail features in Fig.9 (left panel) are not as strongly
defined as in the Cepheid distribution, presumably due to the lack of
dissipation in our collisionless particle model.

The halo component plot in Fig.9, combined with the sky projection
plot of Fig.8 (bottom panel) shows that the halo is roughly spherical
in the
central regions of
the SMC at least within 3 kpc of the centre. A very broad bridge and tail
structure, less well defined compared to the disc simulation,
may be distinguished emerging from the central region.
Although the particle distribution is broad, the locus of the bridge section
tends to smaller distances with decreasing angle defined by CC, whereas the
particles of the tail section are more evenly distributed with respect to
angle.  This is supported by the fact that for negative angles the ratio of the
total number of particles closer than 55 kpc to that beyond 60 kpc is very
nearly unity, whereas the corresponding ratio for positive angles is one-third.
Therefore the model predicts that the depth of the bulk of the stellar
population of the SMC halo would increase and that the mean distance would
decrease moving from SW to NE.   In the studies of the geometry of the SMC
involving the HB/clump  stellar populations in the outer parts of the SMC
(see Hatzidimitriou \& Hawkins 1989,
Gardiner \& Hawkins 1991, and Gardiner, Hatzidimitriou \& Hawkins 1992),
the authors found evidence for large depths of up to $\sim20$ kpc along the
line-of-sight in the eastern part of the SMC, and smaller depths below
10 kpc
in the western areas. Furthermore, they reported a corresponding trend towards
smaller mean distance moduli in the NE in agreement with the model.
A two-component model of the SMC was suggested
by these authors in which a nearer component was superimposed on a more
distant component in the eastern regions. It is seen from Fig.9 that
our best model reinforces such a picture for the eastern region, with
the tidal bridge and tail corresponding to the nearer and more distant
components, respectively. On the other hand, the SW part is populated
mainly by the tail stars, giving rise to a relatively small depth.

\vspace{1cm}
\begin{center}
------

Fig. 9

------
\end{center}
\vspace{1cm}

\noindent
{\em c) Internal kinematics}

In Fig.10 we have plotted the
heliocentric velocity of the particles
with $-78^\circ <Dec< -68^\circ$
against right ascension
for both disc and halo components in order
to investigate the kinematical structure of the SMC.
Firstly, we discuss the disc component (top panel) in relation to a series
of figures produced by Mathewson, Ford \& Visvanathan (1988) (MFV88)
(their Fig.6a-c),
which show contours of H~{\sc i } brightness in the velocity-right ascension
plane for small (40 arcmin diameter) fields in the central parts of the SMC.
In MFV88's Fig.6a and b, which are for regions located in the main bar of
the SMC, there is a strong vertical feature at around $RA=0^h 50^m$.
It is delineated by both the gas and young stars and presumably
corresponds to the velocity field of the bar. The velocity range
of this feature is from about 80-200 kms$^{-1}$ and it is slightly
tilted such that
the eastern part exhibits a higher velocity. This feature corresponds
to the large concentration of particles located at similar right ascension in
the disc
component plot (Fig.10), with a velocity range from 70-220 kms$^{-1}$.
The simulated feature is also tilted in the velocity-right ascension plane
in
the same sense as the observations. The steep gradient of this feature,
which is in excess of 100 kms$^{-1}$/kpc (assuming 1 degree represents
approximately
1 kpc  at the distance of the SMC) is due to the bar being oriented nearly
along the line-of-sight direction.  A large velocity gradient will be
observed when the bar is seen almost end-on because of the highly
non-circular
motions along the bar. The tilt in the velocity-right ascension plane is
due to the specific geometry of the situation in which
larger numbers of approaching (receding) stars are observed
along a given line-of sight on the western (eastern) side
of the bar given that the
particles are rotating clockwise about the SMC centre
in the angle-distance plot of Fig.9.
The H~{\sc i} distribution in MFV88's Fig.6 (especially b) and c))
shows two distinct velocity components in the main body.
This bimodal velocity distribution was not reproduced in the present
model, which may be due to our neglect of dissipation and pressure
effects inherent in the gas component.

Turning now to the velocity structure of the halo component, we see in
Fig.10 (bottom panel) that the velocity feature at around $RA=0^h 50^m$ is
a little broader in right  ascension with a far smaller
 tilt than for the disc, indicating the existence
of a spherical system supported primarily by random motions
in the central regions. This appears to
be consistent with the studies of carbon stars by Hardy, Suntzeff
\& Azzopardi (1989) and planetary nebulae by Dopita {\it et al.} (1985)
which found no evidence of rotation in the central system.
In addition to the bar-induced velocity structure in the disc component, we
also see the velocity components
associated with the tidal bridge and tail in Fig.10 (top panel).
The tail protrudes westward from the main bar at a heliocentric
radial velocity of $\sim 120$ kms$^{-1}$, and then turns to the east
passing the main bar with increasing velocity up to more than
$300$ kms$^{-1}$. The bridge seems to start at a velocity of
$\sim 170$ kms$^{-1}$ eastward from the main bar. It should be noted that
this
latter velocity feature and the start of the tail feature at a velocity
of $\sim 120$ kms$^{-1}$ may have observational counterparts in H~{\sc i}
features seen at similar velocities in MFV88's Fig.6.
In all of their Figs.6a-c, we see a velocity component at
$\sim 180$ kms$^{-1}$ extending eastward from the main component, which can be
identified with the beginning part of the bridge in our model.
Another component at $\sim 120$ kms$^{-1}$ extending westward from the
main bar should correspond to the start of the tidal tail in our model.
Turning to the halo component plot
(Fig.10 bottom panel) it can be seen that the bridge and tail features
are broader in velocity than for the disc component and that the tail
is relatively weaker. It will be demonstrated later that the extension of the
bridge and tail
features of the disc and halo components into the ICR
may help to provide a good explanation of the complicated kinematics of this
area.

\vspace{1cm}
\begin{center}
------

Fig.10

------
\end{center}
\vspace{1cm}

\noindent
{\em d) Velocity-distance correlation}

MFV88 and Hatzidimitriou, Cannon \& Hawkins (1993) (HCH) have reported the
existence
of correlations between the distances and velocities of stars in their
observed samples. MFV88 found a linear relation between the velocities
and distances of some Cepheids observed in the central region and northern
part of the bar (see their Fig.12). They found a slope of 4 kms$^{-1}$/kpc  for
their fit to the data. Hatzidimitriou and her co-workers studied a
sample
of horizontal-branch/clump stars
located at about 3.3 kpc north-east of the SMC optical centre
representing an older stellar group than
the Cepheids, and also found a linear correlation between distance
and velocity but with a larger slope of 8.1 km$^{-1}$/kpc. In order to see
if our best model can provide an explanation for these correlations we have
constructed plots of distance against heliocentric velocity for
disc and halo particles lying
within  $0^h<RA<2^h$, $-78^\circ<Dec<-68^\circ$
which excludes the inter-Cloud and Magellanic Stream areas (see Fig.11).

It can be seen from Fig.11 that the disc and halo components exhibit a
number of common features in the distance-velocity plane. The overall
pattern consists of a central feature
(the bar in the case of the disc component)
between distances of 55 and 60 kpc, a tail section at greater distances
and a bridge section at smaller distances. The bridge and tail sections, which
 are associated with the bridge and tail structures identified in Fig.5,
show correlations of increasing velocity with distance. The correlation is less
 clearly defined for the bridge section than the tail section,
but the general trend is nevertheless apparent.
For the halo component, there is a larger velocity spread at a
given distance
than for the disc, reflecting the greater random motions present in the halo.

We now examine the velocity-distance correlation found by MFV88 for their
sample of Cepheids, which we presume to be associated with the disc
component.
In Fig.11 (top panel), we have superimposed the regression line
representing
the linear correlation of  4 kms$^{-1}$/kpc found to fit the Cepheid
observations. We see a qualitative agreement between this line and the tail
section of the disc component. Since the Cepheids lie mostly between
distances of 55 and 70 kpc (see Fig.12 of MFV88) we conclude that
they are likely to be associated mainly with the SMC's tidal
tail structure (see Fig.9).

To investigate the velocity-distance correlation observed by HCH for
HB/clump stars we consider the halo component plot (Fig.11 bottom panel), since
 these stars belong to the intermediate-aged population in the SMC.
On this plot we have superimposed the fit obtained by HCH for the stellar
observations, which we have indicated by the solid line. Although the
particle distribution is broad, the observational regression line of
8.1 kms$^{-1}$/kpc roughly matches the distance-velocity trend in the bridge
section. A steeper slope in the distance-velocity plane, corresponding
to a gradient closer to the value of 4 kms$^{-1}$/kpc obtained for the
Cepheid sample of MFV88, would appear to give a better representation
of the
particle trend. In fact, a linear fit to the simulation data for particles
between 40 and 55 kpc gives a correlation of 3.4 kms$^{-1}$/kpc.
We note here that HCH's stellar sample  consisted only of stars with
distances less than 60 kpc, which was caused by
a selection effect due to the rejection of more distant stars  with low
signal-to-noise ratios in
the original sample. The stars in HCH's sample therefore appear to
lie mostly in the bridge feature. A rather stronger velocity-distance
correlation for the halo component is seen in the tail section at distances
greater than 60 kpc.
It would certainly be interesting to see if HB/clump stars at larger
distances (for which
HCH failed to get data of sufficient quality during their observing run)
match the velocity-distance
correlation seen in the simulated tidal tail.

On the basis of the existence of velocity-distance trends in the best model
which may be approximately fit by gradients similar to observed
values, we suggest that the origin of the velocity-distance
correlations is connected with the tidal distortion of the
SMC induced by the last close encounter between the Magellanic Clouds about
0.2 Gyr ago.\\

\vspace{1cm}
\begin{center}
------

Fig.11

------
\end{center}
\vspace{1cm}

\noindent
{\bf 4.4 \hspace{0.5cm} The Inter-Cloud Region --- Kinematics}

We now discuss aspects of the simulation of the disc and halo components
for our best
model relating to the kinematics of gas and stars in the region between
the Magellanic Clouds, namely the inter-Cloud region.
Kunkel, Demers \& Irwin (1994) (KDI) have investigated the kinematics of carbon
 stars in the ICR and compared their observations to those
of early-type stars and neutral hydrogen.
They also conducted numerical simulations to investigate the observed
velocity trends. Their Fig.5 shows the galactocentric radial velocities
($V_{GSR}$) of carbon stars,
early-type stars and H~{\sc i} peaks plotted against their angular distance
from the SMC centre. In order to compare these observations to our best
simulation we found it instructive to construct a similar plot for the
disc and halo simulation data (Fig.12). We made plots for the disc
(top panel)
and halo (bottom panel)
particles within the quadrant
centered on the
position angle of the LMC with respect to the SMC centre
to exactly
correspond to the area of sky included
in KDI's figure.

In KDI's Fig.5 the H~{\sc i} velocity peaks
are seen to span a wide range
in galactocentric velocity from $-20$ to 90 kms$^{-1}$ at 5 degrees
from the SMC centre
to a range of 0 to over 150 kms$^{-1}$ at 15 degrees.
Our disc component shows a distribution in velocity which is
a little broader, from -50/130 kms$^{-1}$ at 5 degrees to
-50/200 kms$^{-1}$ at 15 degrees from the SMC centre, but which agrees
remarkably
well with the general trend in the velocity pattern.
KDI noted that the majority of carbon stars   were found at
negative or low galactocentric velocities, whereas it can be seen from their
Fig.5
that the early-type stars are more
evenly distributed within the velocity range defined by the H~{\sc i}
envelope. KDI performed some interpretive numerical simulations which
produced a bridge and tail structure for the SMC.
As already mentioned in Section 4.1, these features appear to be related to the
tidal bridge and tail structure in
our best model. On the basis of their simulation they suggested that
the carbon stars belonged mainly to the bridge section, while the
H~{\sc i} gas belonged mainly to the tail section, some mechanism being
proposed that removes or ionizes the neutral gas in the bridge in order
to account for the observed deficiency there. In our scenario, however,
we contend that the H~{\sc i} gas is associated with both the bridge
and tail sections, not just the tail, and we explain
the difference in the velocity distribution of young (H~{\sc i} gas and
early-type
stars) and old (carbon stars) population components
based on the separate kinematics of the disc and halo components.

Our explanation for the distribution
of carbon star velocities compared to that of the younger populations is as
follows.
We have previously stated that the ICR is populated by the particles of the
tidal bridge and tail. Comparing the disc and
halo components plotted in Fig.12, we can clearly identify the velocity pattern
 due to the bridge and tail at lower and higher galactocentric velocities
respectively in the disc component plot, while in the halo component plot
the bridge is apparent but the tail is very weak. By counting
particles at angular distances between 3.5 and 15 degrees from the SMC
centre,
and ascribing particles with velocity less than 50 kms$^{-1}$ to the
bridge and those with greater velocity to the tail, we could determine the
relative numbers of particles in the bridge and tail for the disc
and halo components. It was found that the bridge/tail particle number ratio
was 1.4 for the disc and 3.3 for the halo.
Confining the distance range to 5-10 degrees leads to corresponding ratios
of 1.8 and 3.6. Thus it is apparent that
the bridge is dominant in the halo component, and therefore our model
predicts
that the carbon stars, which are probably associated with the
halo population,
are more likely to be found in the bridge, while the early-type stars
(and HI gas),
presumably associated with the disc component, are likely to be found in
both
the bridge and tail sections, in good agreement with the observations.
Although the halo component did generate a tidal tail (see Fig.5b)
the tail is not as well developed as the tidal tail from the disc component
in the inter-Cloud region.\\

\vspace{1cm}
\begin{center}
------

Fig.12

------
\end{center}
\vspace{1cm}

\noindent
{\large\bf 5 \hspace{0.5cm} CONCLUSIONS AND FUTURE WORK}

We have carried out an extensive set of numerical simulations
of the tidal distortion of the SMC due to the Galaxy and LMC.
A barred galaxy model was taken to describe the SMC
at the beginning of the interaction simulation well before
the close encounters with the Galaxy and LMC take place.
The effect of particle self-gravity was taken into account for the SMC
model.
Our study is not meant to be a complete survey of the whole permissable
parameter space. However, by proceeding rather heuristically,
we succeeded in obtaining a model which explains many observational
characteristics of the Magellanic system.
The major achievement of the present model is that it has been able to provide
the most coherent explanations to date of various structural and kinematical
properties of the SMC and related tidal features without resorting
to other non-tidal effects such as ram pressure or
collisions with high velocity clouds.

The main results are summarised as follows.\\
1. Our best model has succeeded in reproducing the
observed morphology of the Magellanic Stream
including the general form of the variation of the width of the Stream
from its origin to its tip.
The velocity profile of the H~{\sc i} gas was also reasonably well matched by
the simulation. \\
2. We provide support for the idea of a leading arm on the opposite side of the
Magellanic Clouds to the Magellanic Stream, by identifying
a feature in our simulation which corresponds to the scattered
H~{\sc i} clumps at $260^\circ <l<310^\circ,
-30^\circ<b<30^\circ$ observed by Mathewson, Cleary \&
Murray (1974). \\
3. We have achieved a fairly good reproduction of the observed morphology
of
the SMC bar on the sky plane and its spatial orientation along the
line-of-sight, although the full extent of the bar and narrow spiral arms
observed for Cepheids by Caldwell \& Coulson (1986)
were not reproduced. The underlying kinematical structure of the
central regions was shown to be caused by the velocity pattern induced by
a bar seen almost end-on.

The following three results are based on a detailed examination
of the tidal bridge and tail structure which was formed
as a result of the close encounter between the SMC and LMC
$\sim 2 \times 10^8$ years ago. The outer extension of the tail and bridge
constitute the inter-Cloud region.

\noindent
4. The halo component in our model appears to show a trend of
increasing depth from SW to NE in agreement with studies of the 3D
distribution of horizontal-branch/clump stars in
the outer parts of the SMC (i.e., 2-5 kpc from the optical
centre) by Hatzidimitriou \& Hawkins (1989).
This is because the tidal tail and bridge are both seen superposed
in the NE region on the sky, whereas the SW region
predominantly comprises stars which belong to the tail.\\
5. Our model has succeeded in reproducing the correlation between
distance and velocity found for samples of Cepheid stars observed by
Mathewson, Ford \& Visvanathan (1988) and horizontal-branch/clump stars
observed by Hatzidimitriou, Cannon \& Hawkins (1993).
The Cepheid sample is probably associated with the tidal tail, and the
HB/clump star sample mainly with the tidal
bridge.\\
6. For the inter-Cloud region, the velocity pattern observed
for young objects (neutral hydrogen, early-type stars) and
older objects (carbon stars) showed much
correspondence with the simulated velocity pattern for the disc and halo
components, respectively.
The disc component developed both a tidal tail and a bridge, providing
a natural explanation of the wide velocity range observed for the
neutral gas and the early type stars. On the other hand, the halo mainly
developed a tidal bridge with
the tail being significantly weaker. This may explain
the concentration of carbon stars at low galactocentric
velocities (which correspond to those of the bridge)
found by Kunkel, Demers \& Irwin (1994).
\vspace{0.5cm}

Although the present numerical study succeeded in explaining
many observed structural and kinematical peculiarities of the SMC
and related features within a purely gravitational framework,
we still note several discrepancies.
These discrepancies are mainly related to the young stellar objects and
the gaseous component of the SMC.
For example, the existence of two main velocity components in
the neutral hydrogen distribution of the main body of
the SMC (Mathewson \& Ford 1984) was   not reproduced.
We also failed to simulate the narrow concentration of gas in the
inter-Cloud region which forms the gaseous belt between the
Magellanic Clouds. In addition,
the observed spatial distribution of the Cepheids (Caldwell \&
Coulson 1986) delineates much sharper structures
and the bar is much greater in extent than the model suggests.
It will be quite interesting to see if an adequate gas-dynamical model which
incorporates the dissipative and collisional nature of the interstellar gas
as well as associated star formation processes can solve these problems.
We are now preparing such a study using a cloud-particle scheme
for the interstellar gas model.

Another promising line of future study is to simulate the
dynamics of the LMC (instead of the SMC)
using our self-gravity numerical code.
Although material from the LMC is considered to make a smaller
contribution to the formation of the Magellanic Stream and the
inter-Cloud region,
the internal dynamics of the LMC itself are quite intriguing.
The interstellar gas shows a non-symmetrical spatial distribution
around the LMC bar, and two velocity components are also observed
(the L- and disc- components of Luks \& Rohlfs 1992).
It will therefore be a major goal of
a future study to explain these peculiarities in the LMC
as the outcome of the Galaxy-LMC-SMC interaction.
\vspace{0.5cm}

\noindent{\bf ACKNOWLEDGMENTS}

The authors would like to thank T. Sawa for providing them with
the contour map of neutral hydrogen (the top panel of Fig.8),
and N. Nakazato for preparing Fig.1.

\vspace{1cm}
\noindent
{\bf REFERENCES}

\noindent
Azzopardi M., Rebeirot E., 1991, in Haynes, R. Milne, D., eds, Proc. IAU Symp.
148, The Magellanic Clouds. Kluwer, Dordrecht, p.71\\
Barnes J.E., Hut P., 1986, Nature, 324, 446\\
Caldwell J.A.R., Coulson, I.M., 1986, MNRAS, 218, 223 (CC)\\
Demers S., Irwin M.J., Kunkel W.E., 1993, MNRAS, 260, 103\\
de Vaucouleurs G., 1960, ApJ, 131, 265\\
Dopita M.A., Ford H.C., Lawrence C.J., Webster B.L., 1985, ApJ, 296, 390\\
Faber S.M., Lin D.N.C., 1983, ApJ, 266, L17\\
Feast M.W., Walker A.R., 1987, ARA\&A, 25, 345\\
Fall S.M., 1980, in  Fall S.M., Lynden-Bell D., eds, The Structure and
Evolution of Normal Galaxies. Cambridge Univ. Press, Cambridge, p.1\\
Florsch A., Marcout J., Fleck E., 1981, A\&A, 96, 158\\
Fujimoto M., Sofue Y., 1976, A\&A, 47, 263\\
Fujimoto M., Sofue Y., 1977, A\&A, 61, 199\\
Gardiner L.T., Hatzidimitriou D., 1992, MNRAS, 257, 195\\
Gardiner L.T., Hatzidimitriou D., Hawkins M.R.S., 1992, in MacGillivray H.T.,
Thomson E.B., eds, Proceedings
from the Conference ``Digital Optical Sky Surveys'', Kluwer, Dordrecht,
p.281\\
Gardiner L.T., Hawkins M.R.S., 1991, MNRAS, 251, 174\\
Gardiner L.T., Sawa T., Fujimoto M., 1994, MNRAS, 266, 567 (Paper I)\\
Hardy E., Suntzeff N.B., Azzopardi M., 1989,  ApJ, 344, 210\\
Hatzidimitriou D., Cannon R.D., Hawkins M.R.S., 1993, MNRAS, 261, 873 (HCH)\\
 Hatzidimitriou D., Hawkins M.R.S., 1989, MNRAS, 241, 667\\
Irwin M.J., 1991, in Haynes, R. Milne, D., eds, Proc. IAU Symp. 148, The
Magellanic Clouds. Kluwer, Dordrecht, p.453\\
Jones B.F., Klemola A., Lin D.N.C., 1994, AJ, 107, 1333\\
Kunkel W.E., Demers S., Irwin M.J., 1994,  Proceedings from the
CTIO-ESO workshop on ``The Local Group'', La Serena, Chile (KDI)\\
Laney C.D., Stobie R.S., 1986, MNRAS, 222, 44\\
Lin D.N.C., Jones B.F., Klemola A., 1995, ApJ, 439, 652\\
Lin D.N.C., Lynden-Bell D., 1982, MNRAS, 198, 707\\
Lin D.N.C., Richer H.B., 1992, ApJ, 388, L57\\
Luks Th., Rohlfs K., 1992, A\&A, 263, 41\\
Mathewson D.S., Cleary M.N., Murray J.D., 1974, ApJ, 190, 291\\
Mathewson D.S., Ford, V.L., 1984, in van den Bergh S., de Boer K.S., eds,
Proc. IAU Symp. 108, Structure and Evolution of the Magellanic Clouds. Reidel,
Dordrecht, p.125\\
Mathewson D.S., Ford V.L., Visvanathan N., 1988, ApJ, 333, 617 (MFV88)\\
Moore B., Davis M., 1994, MNRAS, 270, 209\\
Murai T., Fujimoto M., 1980, PASJ, 32, 581 (MF) \\
Ostriker J.P., Peebles P.J.E., 1973, ApJ, 186, 467\\
Rubin V.C., Burstein D., Ford W.K., Thonnard N., 1985, ApJ, 289, 81\\
Schommer R.A., Olszewski E.W., Suntzeff N.B., Harris H.C., 1992, AJ, 103, 447\\
 Sellwood J.A., 1981, A\&A, 99, 362\\
Sofue Y., 1994, PASJ, 46, 431\\
Tanaka K.I., 1981, PASJ, 33, 247\\
Toomre A., 1964, ApJ, 178, 1217\\
Tully B.R., 1988,  Nearby Galaxies Catalog, Cambridge
Univ. Press, Cambridge, p.11\\
van der Kruit P.C., Searle, L., 1982, A\&A, 110, 61\\
Wannier P., Wrixon G.T., 1972, ApJ, 173, L119\\
Westerlund B.E., 1990, A\&AR, 2, 27\\

\newpage

\noindent
{\large\bf  FIGURE CAPTIONS}

\noindent
{\bf Figure 1.}
Schematic view of the Galaxy and the Magellanic Clouds, showing the
relationship between the standard galactocentric coordinate system
($X, Y, Z$)
and the non-inertial SMC-centric system ($x,y,z$)
used for the computations.
The relationship between the two coordinate systems is specified by the
angles $\theta$ and $\phi$ defined by the rotation axis of the SMC
with respect to the galactocentric system.

\noindent
{\bf Figure 2.}
Equilibrium model of the SMC. Projections of the particle positions
on the $x-y$ and $x-z$ planes
of the SMC-centric coordinate system
for both disc and halo components are shown.
Distance units are indicated in kpc.

\noindent
{\bf Figure 3.}
SMC-centred `orbits' of the Galaxy and the LMC for simulations with
$\theta=45^\circ$, $\phi=230^\circ$. The paths traced
by the Galaxy (dashed curve) and the LMC (solid curve)
from $T=-2$ Gyr  to the present are indicated
for SMC-centric $x-y$ ($x$ abcissa), $x-z$ ($x$ abcissa) and
$y-z$ ($y$ abcissa)
projections, with distance units in kpc. The direction of motion is shown
with arrows
as well as the positions at which the LMC-SMC distance is at local
minima at $T=-1.40$ Gyr (filled circle) and $T=-0.18$ Gyr (open circle).
The SMC initially rotates counter-clockwise in the $x-y$ plane.
The bottom right hand plot shows the Galaxy-SMC and LMC-SMC distances
as a function of time.

\noindent
{\bf Figure 4.}
The tidal force exerted on the SMC by the Galaxy and the LMC
as a function of time.
The tidal force was calculated from the double derivative of the Galaxy
and LMC potentials at the position of the SMC (see text for explanation).

\noindent
{\bf Figure 5.}
Global view of the best model at $T=0$ (the present epoch). The position of the
LMC is indicated  by an open circle.\\
(a) Disc component. The disc particles are shown projected onto the SMC-centric
$x-y$ plane with distance units in kpc. Major structures are labelled.\\
(b) Halo component. The halo particles are shown projected onto the SMC-centric
$x-y$ plane with distance units in kpc.

\noindent
{\bf Figure 6.}
Simulation of the Magellanic Stream.
The neutral hydrogen distribution (left panel) and
the distribution of disc particles in the best model (right panel)
are shown projected onto the sky plane. Observational data for the
Magellanic
Stream is taken from Mathewson \& Ford (1984), and the H~{\sc i} clumps
observed by Mathewson, Cleary \& Murray (1974) are also shown indicated by the
thicker curves.

\noindent
{\bf Figure 7.}
The velocity profile of the Magellanic Stream. The variation of the
GSR (galactic standard of rest) velocities of the disc particles in the
best model is shown as a function of Magellanic longitude defined by
Wannier \& Wrixon (1972). Also shown are the observational data of
Mathewson, Cleary \& Murray (1974) represented by diamonds.

\noindent
{\bf Figure 8.}
Comparison of the neutral hydrogen distribution in the vicinity of the
Magellanic Clouds with the particle distribution of the best model.
Top panel: the surface density of H~{\sc i} from Fig.2 of Mathewson \& Ford
(1984).
The contour levels correspond to 5, 10, 20, 40, 75, 150, 400
in units of $10^{19}$ atoms cm$^{-2}$.
This map is meant to roughly illustrate the neutral hydrogen distribution
near the LMC and SMC, and in the inter-Cloud region, so that
some small clumps at the periphery are omitted.
Middle panel: the distribution of disc particles in the best model
plotted on the plane of the sky. Bottom panel: the halo particle distribution.

\noindent
{\bf Figure 9.}
The 3D structure of the SMC. The heliocentric distances of the disc
(left panel)
and halo (right panel) particles of the best model are plotted against the
angular distance along the direction of maximum distance gradient
found by Caldwell \& Coulson (1986), which is in position angle $58^\circ$ and
runs approximately from NE to SW with increasing angle. The distances
of Cepheid variables from this study are also plotted in the left panel
(open circles with error bars).

\noindent
{\bf Figure 10.}
The velocity pattern in the SMC. The heliocentric particle
velocities of the best model are plotted against right ascension for
the disc (top panel) and halo (bottom panel) components.
Only the particles with $-78^\circ<Dec<-68^\circ$
are plotted.
Compare with Fig.6 of Mathewson, Ford \& Visvanathan (1988).

\noindent
{\bf Figure 11.}
The velocity-distance correlation in the SMC.  The heliocentric
distances of particles of the best model lying within $0^h<RA<2^h$,
$-78^\circ<Dec<-68^\circ$, are plotted against their heliocentric velocities
for
the disc (top panel) and halo (bottom panel) components.
The fit representing
the velocity-distance correlation, with a slope of 4 kms$^{-1}$/kpc,
found for a sample of Cepheids by
Mathewson, Ford \& Visvanathan (1988), is indicated in the top panel.
The fit representing the velocity-distance correlation,
with a slope of 8.1 kms$^{-1}$/kpc, found for a sample of
horizontal-branch/clump stars by Hatzidimitriou {\it et al.} (1993),
is indicated in the bottom panel.

\noindent
{\bf Figure 12.}
The velocity pattern in the inter-Cloud region.  The galactocentric
radial
velocities of particles of the the best model lying within
the quadrant centred on the position angle of the LMC are plotted against
angular
distance from the SMC centre for the disc (top panel) and halo (bottom panel)
components.
Compare with Fig.5 of Kunkel, Demers \& Irwin (1994).

\end{document}